\documentclass[prd,twocolumn,showpacs,preprintnumbers,amsmath,amssymb]{revtex4}
\usepackage{graphicx}
\preprint{INT-PUB-09-017}

\begin{document}
\def\gf{\mathfrak{B} }
\def\tc{\lambda^t}
\title{Dyson instability for $2D$ nonlinear $O(N)$ sigma models}
\author{Y. Meurice}
\email[]{yannick-meurice@uiowa.edu}
\affiliation{Department of Physics and Astronomy\\ The University of Iowa\\
Iowa City, Iowa 52242, USA }
\date{\today}
\begin{abstract}
For lattice models with compact field  integration (nonlinear sigma models over compact manifolds and gauge theories with compact groups) and satisfying some discrete symmetry, the change of sign of the bare coupling $g_0^2$ at zero results in a mere discontinuity 
in the average energy rather than the catastrophic instability occurring 
in theories with integration over  arbitrarily large fields. This indicates that the large order of perturbative series and the non-perturbative contributions should have unexpected features. 
Using the large-$N$ limit of 2-dimensional nonlinear $O(N)$ sigma model, we discuss the complex singularities of the average energy for 
complex 't Hooft coupling $\tc= g_0^2N$. A striking difference with the usual situation is the absence of cut along the negative real axis. 
We show that the zeros of the partition function can only be inside 
a clover shape region of the complex $\tc$ plane.
We calculate the density of states and use the result to verify numerically the statement about the zeros. We propose dispersive representations of the derivatives of the average energy for an approximate expression of the discontinuity. The discontinuity is purely non-perturbative and contributions at small negative coupling 
in one dispersive representation are essential to guarantee that 
the derivatives become exponentially small when $\tc\rightarrow 0^+$.
We discuss the implications for gauge theories. 
 \end{abstract}
\pacs{11.15.-q, 11.15.Ha, 11.15.Me, 12.38.Cy}
\maketitle

\section{Introduction}

The lattice formulation of quantum chromodynamics provides a 
non-perturbative ultraviolet regularization and is a widely accepted model for strong interactions. For pure gauge $SU(2)$ and $SU(3)$ with a Wilson action, the absence of phase transition for any real and positive values of $g_0^2$ suggests that it is possible to use a weak coupling expansion in the bare coupling $g_0^2$, which is valid a short distance, to describe the large distance behavior of the theory. 
However, possible limitations of the validity of the weak coupling expansion 
have been raised by Dyson \cite{dyson52}. He argued that if in quantum electrodynamics $e^2$ is changed into $-e^2$, the 
vacuum becomes unstable and that consequently the radius of convergence of the expansion in powers of $e^2$ should be zero. In that article, Dyson says that ``the argument [...] is lacking in mathematical rigor [...] it is intended [...] to serve as a basis for further discussions". The idea was followed beneficially by Bender and Wu \cite{bender69, bender73} and many others \cite {leguillou90} to show that 
the large order behavior of perturbative series can be estimated 
by semi-classical calculations at small negative coupling. 

The factorial growth of perturbative coefficients is due to the large field contributions to the path-integral \cite{pernice98,convpert}. In lattice gauge theory with compact groups, there is a build-in large field cutoff and the theory is well defined  at negative $g_0^2$. For pure gauge 
theory with a $SU(2N)$  gauge group and lattices with even number of sites in each direction, there is an exact discrete symmetry \cite{gluodyn04} that relates the partition function at opposite values of $g_0^2$. 
This symmetry implies that the average plaquette jumps suddenly 
from 0 to 2 as $g_0^2$ goes from very small positive values to very small negative values. So, as $g_0^2$ changes sign, there is a change in vacuum rather than a loss of vacuum. 

The idea that Dyson argument needs to be revisited for lattice models 
with compact gauge groups is also supported by the analysis of 
the expansion of the average plaquette in power of  $g_0^2$. 
Existing series for $SU(3)$ up to  order 10 \cite{direnzo2000} and 16 \cite{rakow05} suggest a power growth rather than a factorial growth. 
These perturbative series are constructed by adding the tails of integration that are absent because of the compactness of the group. 
In the case of a single plaquette \cite{plaquette}, adding the tail of integration leads to a factorial growth. At finite volume, the studies of Refs. \cite{npp} and \cite{Denbleyker:2008ss} suggest 
 that factorial behavior may show up at an order proportional to the volume. The large order behavior of the 
perturbative series is related to the zeros of the partition function in the complex $1/g_0^2$ plane \cite{third,quasig} (Fisher's  zeros). 
This is clearly a difficult problem and it would be useful to understand the connection for lattice models that have the same features but where calculations are easier. 

In this article, we discuss Dyson instability for nonlinear $O(N)$ lattice sigma models. These models are defined in Sec. \ref{sec:model} where we show that their partition functions have the same property 
as $SU(2N)$ gauge theories under the transformation $g_0^2\rightarrow -g_0^2$. In the large-$N$ limit, we can use a saddle point 
approximation. This provides an equation relating the mass gap to 
the 't Hooft coupling $\tc =g_0^2N$. The complex singularities of this map and its inverse are discussed in Sec. \ref{sec:map} and \ref{sec:invmap} respectively. An important feature is the absence of cut 
along the real negative axis in the $\tc$ plane. 

The average energy and the density of states are calculated in Sec. \ref{sec:density} for large-$N$ and compared with weak and strong coupling expansions. In Sec. \ref{sec:zeros}, we discuss Fisher's zeros. We show that they are related to the poles of the average energy and that they can only 
appear in a clover shape region of the complex $\tc$ plane. We use 
the density of states to numerically verify this statement. 
In Sec. \ref{sec:disp}, dispersive methods are proposed to represent 
the derivatives of the average energy in the limit where $\tc \rightarrow 0^+$. These representations are characterized by large contributions canceling each other. 

The article mentions other questions that would be worth addressing in more detail. One is the meaning of the finite radius of convergence for the linear sigma models. This seems to contradict Dyson's argument 
and may be relevant to understand the question of conformal fixed points suggested by Polyakov \cite{Polyakov:2004br}. The other is the 
volume dependence of the non-perturbative part of the average energy and of the locations of Fisher's zeros. These questions  are 
important for numerical calculations in lattice gauge theory.
\section{The model}
\label{sec:model}

\subsection{Basic definitions} 

In this article, we consider the $O(N)$ nonlinear sigma model on a square lattice.  
Most of the results presented in this section can be formulated for arbitrary dimension $D$ not specified until the next section. We call $V=L^D$ the number of sites. 
The lattice sites are denoted ${\mathbf x}$  and the  scalar fields $\vec{\phi}_{\mathbf x}$ are $N$-dimensional unit vectors. The partition function reads:
\begin{equation}
\label{eq:pf}
Z=C\int \prod _{\mathbf x} d^N\phi_{\mathbf x}\delta(\vec{\phi}_{\mathbf x}.\vec{\phi}_{\mathbf x}-1) {\rm e}^{-(1/g_0^2)E[\{\phi\}]}  \   ,
\end{equation}
with 
\begin{equation}
E[\{\phi\}]=-\sum_{{\mathbf x},{\mathbf e}}(\vec{\phi}_{\mathbf x}.{\vec{\phi} }_{\mathbf x+e}-1) \  ,
\end{equation}
with ${\mathbf e}$ running over the $D$ positively oriented unit lattice vectors. 
We introduce the 't Hooft coupling:
\begin{equation}
\tc\equiv g_0^2N \   ,
\end{equation}
that is kept constant as $N$ becomes large.
Its inverse is denoted 
\begin{equation}
b\equiv 1/\tc
\end{equation}

The volume integration at each site is finite and equal 
to the hypersurface of a $N-1$ dimensional sphere $2\pi^{N/2}/\Gamma(N/2)$.
With the normalization $C=(\Gamma(N/2)/2\pi^{N/2})^V$, the partition function becomes 
1 in the limit $b=0$. This is the analog of having the Haar measure normalized to 1 
in lattice gauge theory.

\subsection{Negative coupling duality}

Unlike the linear sigma model, the partition function is well defined at negative coupling. 
The $\vec{\phi}_{\mathbf x}$ are unit vectors, and consequently, 
$-1\leq\vec{\phi}_{\mathbf x}.{\vec{\phi} }_{\mathbf x+e}\leq 1$. This means that 
the energy (or Euclidean action ) $E$  per link is bounded from above and below. 
For a $D$-dimensional  hypercubic lattice with an even number of sites in each direction and periodic boundary conditions, we have 
\begin{equation}
\label{eq:dual}
Z[-g_0^2]={\rm e}^{2DL^D/g_0^2}Z[g_0^2]
\end{equation}
This can be seen by changing variable $\vec{\phi}\rightarrow -\vec{\phi}$ on sublattices with lattice spacing twice larger and such that they share exactly one site with each link of the original lattice. A similar relation can be proven for  $SU(2N)$ pure gauge theories on even lattices \cite{gluodyn04}. 

The argument extends to compact manifolds (for sigma models) and  to compact groups (for gauge theories) provided that it is possible to transform the integration variable into minus itself without affecting the integration measure. It should also be noted that if $g^2_0\neq 0$, $Z[g_0^2]=0$ implies 
$Z[-g_0^2]=0$. 

\def\ave{\mathcal{E}}
The average energy per unit of volume $V$ and its derivatives provide important information about possible phase transitions.  We denote it 
$\ave\equiv <E>/V$. 
With our notations,
\begin{equation}
\label{eq:avedef}
\ave=-(1/(VN))\partial {\rm ln} Z/\partial b \ .
\end{equation}
The symmetry (\ref{eq:dual}) implies the sum rule
\begin{equation}
\ave(-g_0^2)+\ave(g_0^2)=2D\ .
\end{equation}
Knowing that when $g_0^2\rightarrow 0^+$, $\ave(g_0^2)\rightarrow 0$, the sum rule implies that if $g_0^2\rightarrow 0^-$, then $\ave(g_0^2)\rightarrow 2D$. In other words, there is a discontinuity in the average energy when $g_0^2$ changes sign. 
\subsection{The gap equation}

In the large-$N$ limit, it possible to calculate the partition function in the 
saddle point approximation \cite{Novikov:1984ac, david84, Polyakov:1987ez}.
In the case of the nonlinear sigma model, one 
enforces the condition $\vec{\phi}_{\mathbf x}.\vec{\phi}_{\mathbf x}=1$ using a Lagrange multiplier. The integration over $\phi$ can then be done exactly. Varying the 
zero mode of the Lagrange multiplier, we obtain:
\begin{equation}
\label{eq:gap}
b=\gf(M^2)\ ,
\end{equation}
with
\begin{equation}
\gf(M^2)\equiv 
\prod_{j=1}^D\int_{-\pi}^{\pi}\frac{dk_j}{2\pi}\frac{1}{2(\sum_{j=1}^D(1-{\rm cos}(k_j))+M^2\ .}
\end{equation}
$M^2$ is the saddle point value of the suitably rescaled Lagrange multiplier and can be interpreted as the mass gap or as the renormalized mass in cutoff units. 
At finite volume, the integral is replaced by a sum over momenta $n_i2\pi/L$.
 
The saddle point equation is invariant under the simultaneous changes:
\begin{eqnarray}
\label{eq:sym}
\tc&\rightarrow&-\tc \\ \nonumber
M^2&\rightarrow& -M^2-4D \ .
\end{eqnarray}
This can be seen by changing variables $k_j\rightarrow k_j+\pi$ for all $j$. 
Note that this change of variable sends the zero-momentum mode into the fastest 
oscillating one (that changes sign at every lattice site). 

It is interesting to compare the gap equation (\ref{eq:gap}) with its counterpart for the 
linear sigma model. In the linear case, the Lagrange multiplier is used to replace 
$\vec{\phi}_{\mathbf x}.{\vec{\phi} }_{\mathbf x}$ by a composite field. 
After suitable rescalings \cite{david84}, this composite field is denoted $X$ and the gap equation becomes 
\begin{eqnarray}
X&=&\gf(M^2)\\  \nonumber
M^2&=&2U'(X)
\end{eqnarray}
for a rescaled bare potential $U(X)$. In the case of a $\phi^4$ theory, we can choose 
$U(X)=(m^2_B/2) X+\tilde{\lambda} X^2$, where $m_B^2$ is the bare mass and $\tilde{\lambda}$ is kept constant when $N$ becomes large. 
$X$ can then be eliminated yielding:
\begin{equation}
(M^2-m_B^2)/\tilde{\lambda}=\gf(M^2)\  .
\end{equation}
Note that despite the fact that the partition function is not well defined at negative 
$\tilde{\lambda}$, the $D=2$ gap equation  has 2 real solutions for $M^2$ when  $\tilde{\lambda}_c<\tilde{\lambda}<0$. At $\tilde{\lambda}_c$ the two solutions merge.  They disappear in the complex plane when $\tilde{\lambda}<\tilde{\lambda}_c$. 

The comparison between the linear and nonlinear sigma models shows that there 
are important differences between the two cases. In the linear case, the situation is 
very similar to what is observed in other scalar models \cite{leguillou90} (dispersion 
relations with a cut extending to $-\infty$), but the nonlinear case is quite different as 
we now proceed to explain.

\section{The gap equation at complex coupling for $D=2$}
\label{sec:map}

In this section, we study the map $\gf(M^2)$ when 
$M^2$ is varied in a cut complex plane. 
The cut is between -8 and 0 on the real axis. 
$D=2$ is assumed through this section. From the previous section, we know that 
\begin{eqnarray}
\label{eq:4symm}
\gf(-8-M^2)&=&-\gf(M^2) \nonumber \\  
\gf(M^{2*})&=&\gf(M^2)^*\ ,
\end{eqnarray}
and we only need to study one quadrant, for instance $Re\ M^2>-4$ and $Im\ M^2>0$ .

\subsection{Large $M^2$ behavior}

The basic observation is that for large $|M^2|$, 
we have an approximate one-to-one mapping since $\tc \simeq M^2$. This is fundamentally different from what happens for the linear sigma model where $ \tilde{\lambda}\simeq (1/4)(M^2)^2$ 
and the inverse mapping requires a cut that can be taken along the negative axis.

The approximation can be improved and we can calculate the strong coupling expansion
\begin{equation}
1/\tc=1/M^2-4/M^2+20/M^6\dots
\end{equation}
and its inverse
\begin{equation}
\label{eq:scl}
1/M^2=1/\tc+4/(\tc) ^2+12/(\tc) ^3\dots
\end{equation}
Numerical studies of the large order expansion in powers of $1/M^2$ show  a clear evidence for a singularity at $1/M^2=-1/8$. 

\begin{figure}
\includegraphics[width=3.3in,angle=0]{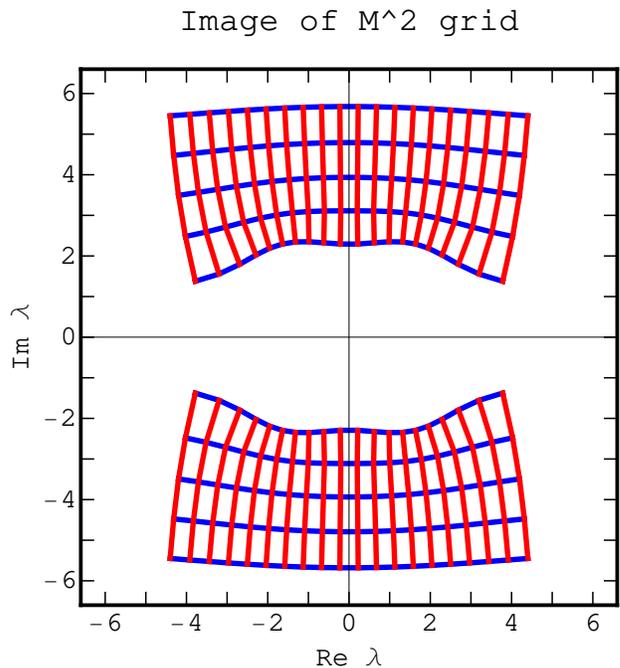}
\caption{
\label{fig:mapl} Complex values of $\tc=1/ \gf(M^2)$  when $M^2$ varies over horizontal (with a spacing  1) and vertical lines (with a spacing 0.5) centered about the cut in the complex $M^2$ plane.}
\end{figure}

As we reduce $|M^2|$, the image of a cartesian grid gets distorted and gaps open near the origin. This illustrated in Fig. \ref{fig:mapl}. The approach of the cut is intricate and 
involves logarithmic divergences in $\gf(M^2)$ that we now proceed to study. 
\subsection{Logarithmic divergences of  $\gf(M^2)$}
\begin{figure}
\includegraphics[width=3.3in,angle=0]{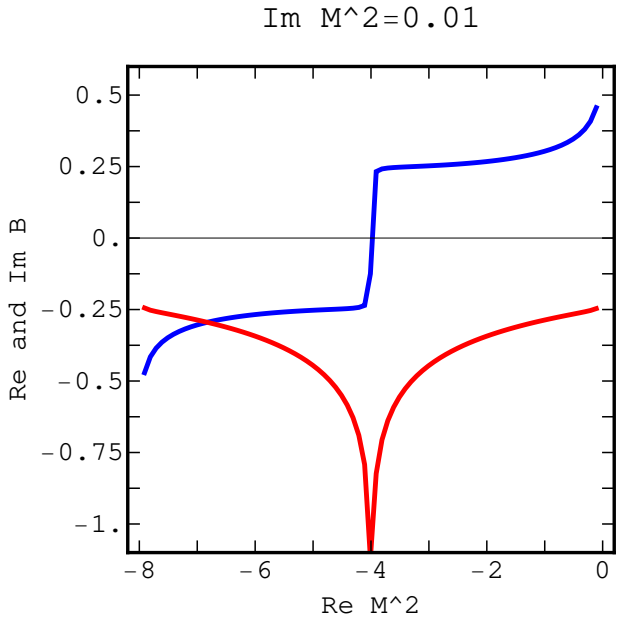}
\caption{
\label{fig:reimb} Real (blue online) and Imaginary (red online) part of $\gf(M^2)$  when $M^2$ varies over a horizontal line 0.01 above the cut in the complex $M^2$ plane. The 
imaginary part has a negative spike at -4.}
\end{figure}

For $D=2$, 
logarithmic divergences appear in $\gf(M^2)$ from region of integration 
where 
\begin{equation}
\label{eq:cont}
2(2-{\rm cos}(k_1)-{\rm cos}(k_2)) \simeq A \pm k_1^2\pm k_2^2\ .
\end{equation} 
This only occurs when $k_i=$ 0 or $\pi$. The four cases  are $A$ = 
0 (for (0,0)), 4 (for (0,$\pi$) or ($\pi$,0)) and 8 (for $(\pi,\pi))$. 
Logarithmic divergences appear when $M^2$ approaches 0, -4 and -8. 
In order to give a first idea, we have plotted in Fig. \ref{fig:reimb} the real and imaginary part of $\gf(M^2)$ 
when $M^2$ runs over a line slightly above the cut.  This figure suggests that the real part has singularities at 0 and -8 and that the imaginary part has singularities at -4. 

The
leading coefficients of the logarithmic singularities can be estimated by using the continuum approximation (\ref{eq:cont}). For instance, if $M^2\rightarrow 0^+$ on the 
real axis, we obtain the familiar relation $\gf(M^2)\simeq -(1/4\pi) {\rm ln}(M^2)$. 
For $M^2\simeq -4$, we have two contributions and by applying the proper 
Wick rotations, we obtain an imaginary expression that is twice larger in absolute 
value than the real part near  for $M^2\simeq 0$. 

\begin{figure}
\includegraphics[width=3.3in,angle=0]{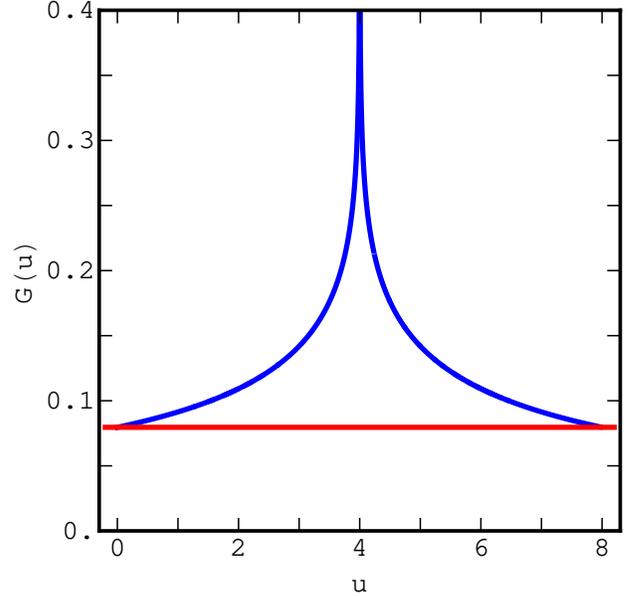}
\caption{{\label{fig:gofu}}$G(u)$ (blue online) compared to $1/(4\pi)$ (red online)}
\end{figure}

Further insight into the singularities can be obtained by introducing the spectral decomposition :
\begin{equation}
\label{eq:decomp}
\gf(M^2)=\int _0^8 du G(u) \frac{1}{u+M^2}\  ,
\end{equation}
with
\begin{eqnarray}
G(u)&\equiv & \int \frac{d^2k}{(2\pi)^2}\delta(u-2(2-\cos(k_1)-\cos(k_2)))\\ \nonumber
&=&\frac{1}{2\pi^2}\int_{-1}^1dc \frac{\theta(1-|u/2-2+c|)}{\sqrt{1-c^2}\sqrt{1-(u/2-2+c)^2}}\ .
\end{eqnarray}
The symmetry (\ref{eq:sym}) implies that 
\begin{equation}
G(8-u)=G(u)\ .
\end{equation}
Numerical values for $G(u)$ are provided in Fig. \ref{fig:gofu}. 
The estimate of the singularity of the real part of $\gf(M^2)$ near $M^2=0$ implies 
\begin{equation}
G(0)=1/4\pi \ ,
\end{equation}
which is confirmed numerically in  Fig. \ref{fig:gofu}. 
From the residue theorem, we find that 
the imaginary part has a discontinuity along the cut. For $M^2 \in [-8,0]$, 
\begin{equation}
 {\rm Im}( \gf(M^2\pm i\epsilon)) = \mp \pi G(-M^2)\ .
\end{equation}
From the analysis of the logarithmic singularity of the imaginary part of $\gf(M^2)$ near $M^2=-4$, this implies that 
for $u\simeq 4$, 
\begin{equation}
G(u)\simeq -(1/(2\pi^2)){\rm ln}( |u-4|) \ .
\end{equation}
This result was confirmed by an analysis of the 
numerical values of $G(u)$ near $u=4$. 
Finally, using the fact that when $u$ reaches 4, two pairs of inverse square root singularities 
coalesce into two poles, it is possible to justify that the real part jumps suddenly from -1/4 to 1/4 
as $M^2$ increases and crosses -4.

By construction, the integration path in Eq. (\ref{eq:decomp}) does not 
wrap around the pole at $u=-M^2$.  As in the case of the logarithm, 
we can introduce 
\begin{equation}
\label{eq:bk}
\gf_k(M^2)\equiv \gf (M^2)+ik2\pi G(-M^2)\ ,
\end{equation}
which corresponds to have the contour of integration wrapping $k$ 
times around the pole. 

\subsection{$\gf$ image of the cut $M^2$ plane}

\begin{figure}
\includegraphics[width=3.3in,angle=0]{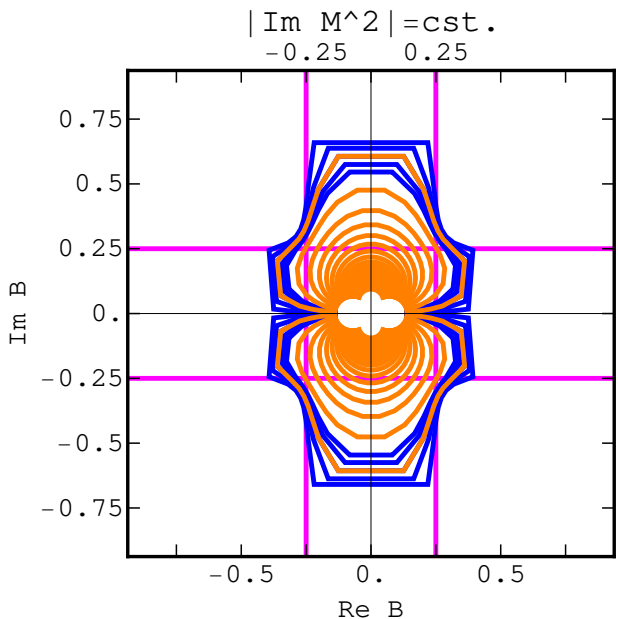}
\caption{\label{fig:bregion}
Complex values taken by $\gf(M^2)$ when $M^2$ varies over the complex plane (here on horizontal lines in the $M^2$ plane with spacing 0.1 (black, blue online) and 0.5 (gray, orange online)). 
Asymptotic limits are $\pm 0.25$ in both direction.}
\end{figure}

We are now in position to determine the image $\gf(M^2)$ of the cut $M^2$ plane. The region 
of large $|M^2|$ is mapped into a neighborhood of the origin. As we approach the cut, the image of 
lines of constant imaginary $M^2$ are mapped into hat-shaped curves shown in Fig. \ref{fig:bregion}. In the limit of zero imaginary part, the curves become approximate hyperbolas with asymptotes on the boundary of a cross of width 0.5 centered at the origin. The asymptotes correspond to the 
logarithmic singularities and can be read from Fig. \ref{fig:reimb}. When $M^2$ has 
zero real part and a small positive imaginary part, the imaginary part of $\gf(M^2)$ reaches -0.25 
while the real part becomes large and positive. The other asymptote of the approximate hyperbola 
is reached by approaching -4 from above with  a positive imaginary part, the real part of $\gf(M^2)$ then reaches 0.25  
while the imaginary part becomes very negative.  All the other cases can be obtained from the symmetry (\ref{eq:4symm}). 
\subsection{$\tc $ image of the cut $M^2$ plane}

\begin{figure}
\includegraphics[width=3.3in,angle=0]{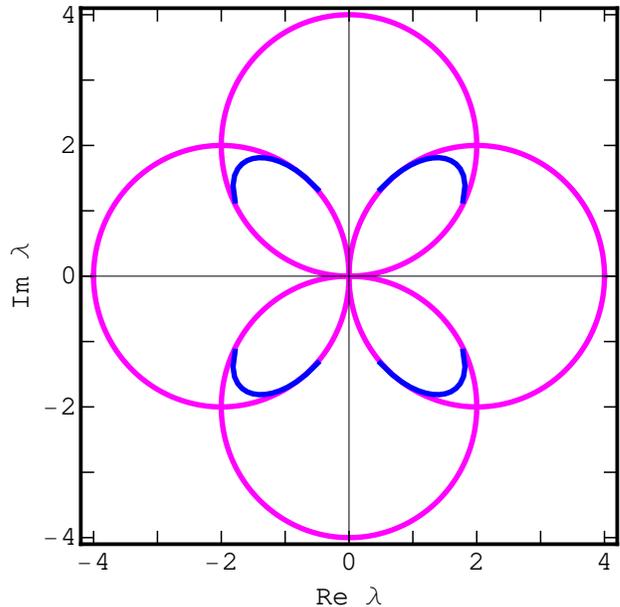}
\caption{
\label{fig:clover} Complex values taken by $\lambda^t$ when $M^2$ varies over lines above and below the cut with $Im M^2=\pm 0.01$ (black, blue online); the circles are the inverses of the asymptotic lines in Fig. \ref{fig:bregion}.}
\end{figure}

We can now draw more precisely the empty region in the middle of Fig. \ref{fig:mapl} . Before constructing it, we state the final result: the image of the cut complex $M^2$ plane 
under the $1/\gf(M^2)$ map 
is the complex plane minus a clover shape centered at the origin and with lobes approximately bisecting the real 
and imaginary axis. The clover shape shape is visible in Fig. \ref{fig:clover}.

The boundary of the clover shape is the $1/z$  map of the 4 limiting approximate hyperbolas discussed in the previous subsection. First, we construct the image of the asymptotes. 
Their images are circles of radius 2 centered at $\pm 2$ and $\pm 2i$. Near the origin in the $\tc$ plane, the circles 
provide a good approximation of the boundary. As we move away from the origin, numerical values 
are necessary. This is illustrated in Fig. \ref{fig:clover}. 

\section{Singularities of $M^2(\tc)$}
\label{sec:invmap}

In Sec. \ref{sec:map}, we have constructed a one-to-one map between the cut $M^2$ plane 
and the $\tc$ plane with a clover removed. The clover can be filled by adding the images $\gf_k$ indexed by an integer $k$  as defined in Eq. (\ref{eq:bk}). 
In this Sec., we construct the inverse mapping and discuss its singularities. We start with a simplified example where everything can be done analytically and then discuss the original problem.
\subsection{A simplified example}

We will start by constructing the inverse mapping for a simplified mapping $\gf_{simpl.}(M^2)$ for 
which $G(u)$ in the decomposition (\ref{eq:decomp}) is constant and equal to $1/4\pi$. This implies 
\begin{equation}
\label{eq:simp}
\gf_{simpl.}(M^2)=(1/4\pi){\rm ln}(1+8/M^2) \ .
\end{equation}
This modification preserves the logarithmic singularities at 0 and -8 and the symmetry 
(\ref{eq:sym}). This example is probably closer to what we expect to find for 
$4D$ lattice gauge theories. The image of the cut complex $M^2$ plane (fundamental domain) is the horizontal strip 
bounded by the horizontal lines with imaginary part $\pm 0.25$.  
The inverse map can be calculated explicitly: 
\begin{equation}
\label{eq:invsimp}
M^2_{simpl.}(b)=8/({\rm e}^{4\pi b}-1) \   ,
\end{equation}
and is invariant under $b\rightarrow b+ ik/2$, for any integer $k$.
The translated values obtained from the fundamental domain, could be obtained directly by having the path of integration in Eq. (\ref{eq:decomp}) wrapping 
$k$ times around $-M^2$.  
The poles of  $M^2_{simpl.}(b)$ are located at $b=0, \ \pm i/2, \ \pm i ,\dots$.

\begin{figure}
\includegraphics[width=3.in,angle=270]{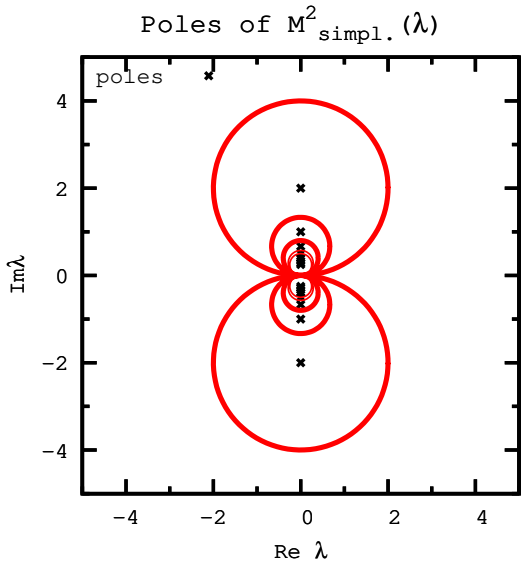}
\caption{\label{fig:simple} A finite number of poles and boundaries of domains described in the text  for $\lambda _{simpl.}(M^2)$ in the $M^2$ plane. }
\end{figure}

In the $\tc$ plane, the fundamental domain is the complex plane with two circles removed . The boundary of the $k$-translated domains 
are circles of radius $2/(2|k|+1)$ and centered at $\pm i2/(2|k|+1)$. 
The poles are located at $-i2/k$.  A finite number of boundaries and poles are shown in Fig. \ref{fig:simple}. For large $|k|$, the poles and 
boundaries accumulate at the origin. 
We believe that some qualitative features of this 
picture are representative of what is encountered in lattice gauge theory. 

\subsection{Qualitative features of $M^2(\tc)$}

We can now describe qualitatively what happens when we restore 
the original features of $G(u)$. The fundamental domain in the $b$ 
complex plane becomes 
strongly distorted when $Re b$ is small as already shown in Fig. \ref{fig:bregion}. The translated domains that were parallel strips in the 
simplified example are obtained by adding $ik2\pi G(-M^2) $ as 
specified by Eq. (\ref{eq:bk}). The reader can visualize the effect by 
combining Figs. \ref{fig:bregion} and \ref{fig:gofu}. 

We can now discuss qualitatively  
the deformation of Fig. \ref{fig:simple} that 
the logarithmic divergence of $G(u)$ near $u=4$ imposes. 
In the $b$ plane the strips are pulled at infinity when $Re b$ becomes 
small. In the $\tc$ plane, the circles are pushed toward the origin along 
the imaginary axis forming concentric clover shape figures. 
Importantly, all the poles have moved to the origin.   
\section{The density of states}
\label{sec:density}

\subsection{Average energy}

From the definition of the average  energy Eq. (\ref{eq:avedef}), we obtain that in the saddle point approximation
\begin{equation}
\label{eq:ave}
\ave=(1/2)(\tc -M^2)\ .
\end{equation}
Note that $0\leq\ave\leq4$  and the range is $N$-independent.

When $\tc$ approaches 0 on the positive real axis, $M^2$ approaches 
zero like  $8{\rm e}^{-4\pi/\lambda}$. We call $-(1/2)M^2$ the non-perturbative part of $\ave$. The perturbative expansion  terminates at first order. For large $\tc$, the leading terms cancel and can use the 
strong coupling expansion (\ref{eq:scl}) to obtain
\begin{equation}
\label{eq:sce}
\ave=2-2/\tc \dots
\end{equation}

\subsection{Saddle point calculation of $n(S)$}

\begin{figure}[b]
\includegraphics[width=3.3in,angle=0]{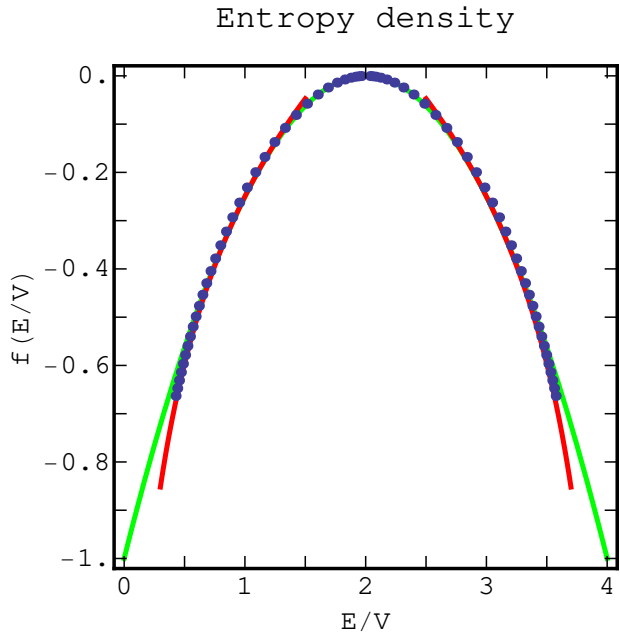}
\caption{
\label{fig:fe} $f(\mathcal{E})$, numerical (circle), first order strong coupling (parabola, green online) and first order weak coupling (red online).}
\end{figure}

We can use a spectral decomposition of the partition function for 
all possible energies: 
\begin{equation}
Z=\int dEn(E){\rm e}^{-bNE}\ ,
\end{equation}
with $n(E)$, the density of state which  can be defined as 
\begin{equation}
n(E)=\int \prod _{\mathbf x} d^N\phi_{\mathbf x}\delta(\vec{\phi}_{\mathbf x}.\vec{\phi}_{\mathbf x}-1)\delta(E[\{\phi\}]-E)
\end{equation}
$n(E)$ is non-zero only if $0\leq E\leq DV$ in $D$ dimensions and this 
implies that $Z$ is an analytical function in the entire $b$ plane. 
\def\lag{\alpha}
Using
\begin{equation}
\delta(E[\{\phi\}]-E)= \int_{K-i\infty}^{K+i\infty}d\lag{\rm e}^{\lag (E[\{\phi\}]-E)}\ .
\end{equation}
and varying with respect to $M^2$ (introduced as before) and $\lag$, we obtain
\begin{eqnarray}
\nonumber
\label{eq:ne}
\lag&=&\gf(1/\lag-2\mathcal{E}) \\
M^2&=&1/\lag-2\mathcal{E} \  ,
\end{eqnarray}
with $\lag$ and $M^2$ understood as  functions of $\ave$. 
In these equations $\ave$ is the independent variable and they are equivalent to Eqs. (\ref{eq:gap}) and 
(\ref{eq:ave}) provided that we identify $\lag$  and $b=1/\tc$. 
From these results, we obtain
\begin{equation}
n(E)\propto{\rm e}^{VNf(E/V)}\ ,
\end{equation}
with the entropy density
\begin{eqnarray}
\label{eq:fe}
& &f(\ave)=-(1/2){\rm log}(\lag)\\ \nonumber
&\ &-(1/2)\prod_{j=1}^2\int_{-\pi}^{\pi}\frac{dk_j}{2\pi}{\rm ln}(2(\sum_{j=1}^2(1-{\rm cos}(k_j))+M^2)
\end{eqnarray}
The duality (\ref{eq:dual}) implies 
\begin{equation}
f(\ave)=f(4-\ave)\ .
\end{equation}
Numerical values of $f(\ave)$ are shown in Fig. \ref{fig:fe}. 
They were calculated using $M^2$ as a parameter first used to fix 
$\lag$ and  $\ave$ from (\ref{eq:ne}) and then $f(\ave)$ from 
Eq. (\ref{eq:fe}).

If we rewrite the partition function in terms of the density of states, 
we obtain the usual thermodynamic relation:
\begin{equation}
\label{eq:beq}
f'(\ave)=b\ .
\end{equation}
One can check that the saddle point equations imply $\lag=b$ as 
expected. 

The behavior of $f(\ave)$ can be approximated near the origin. 
Using $\ave \simeq (1/2)\tc$ and Eq. (\ref{eq:beq}), we obtain 
\begin{equation}
f(\ave)\simeq (1/2){\rm ln}(\ave)\ .
\end{equation}
The behavior of $f(\ave)$ near its maximum at $\ave=2$ can be 
approximated using Eqs. (\ref{eq:sce}) and (\ref{eq:beq}). The result is
\begin{equation}
f(\ave)\simeq (-1/4)(\ave-2)^2
\end{equation}
These two approximations fit the numerical values quite well in their region of validity as shown in Fig. \ref{fig:fe}.

As long as $\ave$ takes values in the real interval $[0, 4]$, Eq. (\ref{eq:beq}) defines a one-to-one mapping between this interval and the 
entire real axis. It is clear from Fig. \ref{fig:fe} that in this interval $f''(\ave)<0$ and consequently the derivative of the mapping is never zero over the interval. This allows us to integrate over the quadratic fluctuations about the minimum. If an analytical continuation of $f(\ave)$ can be constructed, it seems clear that poles and zeros of 
$f''(\ave)$ will play an important role in the determination of the zeros 
of the partition function which is the subject of the next section. 

\section{Zeros of the partition function}
\label{sec:zeros}

There exists a simple relation between the poles of the average energy and the zeros of the partition function. If $b_0$ is a zero of $Z$ 
of order $q$, then  $(dZ/db)/Z\simeq q/(b-b_0)$ for $b\simeq b_0$. 
If we now integrate over a closed contour $C$, 
\begin{equation}
\oint_C  db (dZ/db)/Z =i2\pi \sum_q n_q(C)\ ,
\end{equation}
where $ n_q(C)$ is the number of zeros of order $q$ inside $C$
Using Eq. (\ref{eq:ave}), we obtain that in the large $N$ limit, 
\begin{equation}
(4\pi i)^{-1}\oint_C  db(M^2-1/b) =\sum_q n_q(C)  /(VN)\ .
\end{equation}
In this expression, $M^2$ is understood as a function of $b$ by inverting Eq. (\ref{eq:gap}). 
The second term has a pole at $b=0$, but it is compensated by a pole 
in $M^2$. This is due to the fact the $b\simeq 1/M^2$ for small $|b|$. 
We now consider possible poles of $M^2$ for other values of $b$. 
We change variable to write
\begin{equation}
\oint_C  dbM^2= \oint_{C'}  dM^2 (db/dM^2) M^2 
\end{equation}
where $C'$ is the contour corresponding to $C$ in the $M^2$ plane and $b$ a short notation for $\gf(M^2)$.
At finite volume, it is possible write $\gf(M^2)$ as a ratio of two polynomials. The zeros of the denominator can only be in the cut (the real interval $[-8,0]$). This property persists for arbitrarily large volume. Consequently, if the contour $C'$ in the $M^2$ plane does 
not cross the cut, then there are no zeros of the partition function inside the corresponding $C$ in the $b$-plane.  We conclude that in the large-$N$ limit, there are no Fisher's zero in the $\gf(M^2)$ image of the cut $M^2$ plane. This image has been constructed in Sec. \ref{sec:map} and limited by four approximate hyperbolas with asymptotes 
along a cross shaped figure. In the $\tc$ plane, this region becomes the complement of a clover shape figure (see Figs. \ref{fig:bregion} and \ref{fig:clover}). 
 
\begin{figure}
\includegraphics[width=3.in,angle=0]{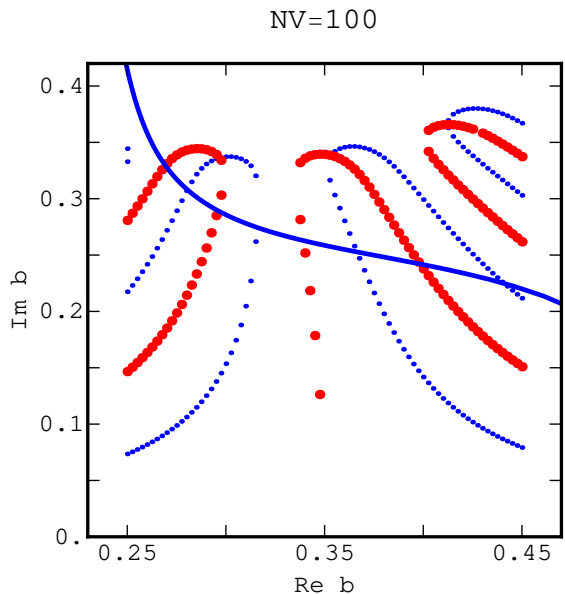}
\caption{\label{fig:zeros}Fisher's zeros for $NV=100$. 
Zeros of $ReZ$ (small dots, blue online), zeros of $ImZ$ (larger dots, red online). 
The solid  line (blue online)  is the image of a horizontal line slightly below the cut in the $M^2$ plane.}
\end{figure}

The argument has been checked by numerical calculations using 
methods similar to those used in lattice gauge theories 
\cite{Denbleyker:2008ss, quasig}. We used 
spline interpolations from 400 numerical values of $f(\ave)$ in Eq. (\ref{eq:fe}) and $NV=100$. 
In order to remove fast oscillations, we have subtracted the average value of the energy at $b=0.35$ from $\ave$ in the exponential. 
This does not affect the complex zeros. 
The results are shown in Fig. \ref{fig:zeros} for a rectangle with 0.1 on each side of $b=0.35$ and $Im b\leq 0.4$ (beyond that new methods need to be developed to calculate the rapidly oscillating integrals). 
The Fisher's complex zeros are the three isolated points at which the zeros of the imaginary part of $Z$ meet those of the real part. The image of a line slightly below the cut ($ImM^2$ =-0.05) is also shown. As predicted there are no Fisher's zeros below the image of this line. 

The construction points out a relationship between Fisher's zeros 
and the poles of $db/dM^2$. This derivative also appears in the 
$\beta$-function
\begin{equation}
\beta(\tc)\equiv\Lambda d \tc/d\Lambda=2(\tc)^{2} M^2db/dM^2\ , 
\end{equation}
assuming $M^2=m^2_R/\Lambda^2$ with the renormalized mass $m_R^2$ kept fixed and $\Lambda$ the UV cutoff. In the approximation where $b=-(4\pi)^{-1}{\rm ln}M^2$, we recover the 
well-known continuum result $\beta(\tc)=-(\tc)^2/(2\pi)$.
Following the finite volume reasoning used above, we see 
that the poles are exactly cancelled by zeros of $(\tc)^{2}$. 
The zeros of the beta function can be interpreted as the 
singular points of the $\tc(M^2)$ map which have nontrivial locations 
at finite volume. 

\section{Toward a dispersive approach}
\label{sec:disp}

In quantum mechanics or linear scalar models with a $\lambda \phi^4$ interactions, it is  common \cite{bender69, bender73,leguillou90} to use dispersion relations to estimate the 
large order behavior of perturbative series. Typically, 
one considers a quantity, that we will denote $F(\lambda)$, that is analytical in the cut plane with a cut going from $-\infty$ to 0. The coefficients $F_k$ of the perturbative series in powers of $\lambda$ are then expressed as 
\begin{equation}
F_k=(1/\pi)\int_{-\infty}^0d\lambda\  Im F(\lambda)/\lambda^{k+1}\ .
\end{equation}
For large $k$, the integral is dominated by small negative values of 
$\lambda$ and one can use semi-classical methods to estimate 
$Im F(\lambda)$ in this regime. Typically, $Im F(\lambda)\sim \lambda^{-b}{\rm e}^{a/\lambda}$ which leads  to a factorial growth $F_k \sim(- a)^{-k}\Gamma(b+k)$.

We are interested in finding a  dispersive representation of $\ave(\tc)$ 
as expressed in the leading order Eq. (\ref{eq:ave}).  In this approximation, the series terminates and what we would expect to learn 
from dispersive  methods is that $M^2$, which appears in the second term of $\ave$, is zero to all order in $\lambda$, when $\tc \rightarrow 0^+$. 

In the rest of this section we will study dispersive expressions for 
$M^2_{simpl.}$ defined in Eq. (\ref{eq:invsimp}). 
This simplified expression of the mass gap does
has a different behavior 
when $\lambda$ approaches 0 along the imaginary axis but it preserves the basic symmetry (\ref{eq:sym}) and it probably  has more  resemblance with the gauge models. 
This choice also has the advantage that all the calculations can be done in terms of elementary 
functions and that the correctness and accuracy of dispersive expressions can be checked easily. 
\def\di{\mathfrak{R} }
We define
\begin{equation}
\label{eq:di}
\di _m(\epsilon)=(i2\pi)^{-1}\oint d\lambda (\lambda-\epsilon)^{-m-1}M^2_{simpl.}(\lambda) \ ,
\end{equation}
where the contour of integration runs counterclockwise along a circle of center $\epsilon$ and radius smaller than $\epsilon$.  In the rest of this section, $\epsilon$ is real and strictly positive. 
By construction, $m!\di _m(\epsilon)$ is the $m-th$ derivative of $M^2_{simpl.}$ evaluated at $\epsilon$. 
If $\epsilon$ is not exactly zero, we can obtain analytical expressions, 
such as 
\begin{equation}
\di _1(\epsilon)=8\pi/({\rm sinh^2}(2\pi/\epsilon) \epsilon ^2 )\  .
\end{equation}
The $\di _m(\epsilon)$ calculated that way are even functions of 
$\epsilon$ for $m$ odd and larger than 1 and odd functions for $m$ even and larger than 2. However, since $M^2_{simpl.}$ has a discontinuity at 0, $\delta(\epsilon)$ and derivatives should also be present if we wish to obtain expressions valid over the whole real axis. 

For $0<\epsilon << 4\pi$, $\di _m(\epsilon)$ can be approximated by 
a sum of terms of the form ${\rm e}^{-4\pi/\epsilon}\epsilon^{-q}$, with 
$q\leq 2m$. These individual terms peak at $\epsilon =4\pi/q$ where 
their value grow like $q!$. When $m$ increases, there is an intermediate region where many large terms cancel and it is difficult 
to evaluate the function numerically. For instance, for $m$ = 12, some 
numerical noise becomes visible for $0.3<\epsilon<0.8$ when 16 digit 
arithmetic is used. 

For $m\geq2$, the contribution of the circle at infinity vanishes, and we can deform the contour in Eq. (\ref{eq:di}) into a clockwise contour encircling the poles on the imaginary axis. 
Due to the absence of cut or poles on negative real axis, the contribution of two line integrals running on opposite directions along 
the negative axis cancel. This is contrast to what happens for quantum mechanics and linear scalar models, where there is a discontinuity.  Using the residue theorem, we obtain
\begin{equation}
\di _m(\epsilon)=-(8/\pi)\sum_{k\neq 0}k^{m-1}(2i-\epsilon k)^{-m-1}\ .
\end{equation}
This expression provides reasonably accurate expressions with truncated sums when $\epsilon$ is not too small. When $\epsilon$ becomes small, the sum can be divided into two parts, one with $|k|<2/\epsilon$ and the other with the rest of the terms. Each sum is of order 
$\epsilon^{-m}$ and the two large sums must cancel to yield an 
exponentially small results.

Another option is to start with the deformed contour encircling the poles on the 
imaginary axis keeping it slightly outside of the figure 8 visible on Fig. 
\ref{fig:simple}. Changing variable to $M^2$, this contour is then mapped into a contour encircling the cut 
from -8 to 0 (but not not containing $M^2_{simpl.}(\epsilon)$). 
With this contour,  we have 
\begin{equation}
\di _m(\epsilon)=(2\pi i)^{-1}\oint dM^2b'b^{m-1}(1-b\epsilon)^{-m-1}\ ,
\end{equation}
with $b$ a short notation for $\gf_{simpl.} (M^2)$ given in 
Eq. (\ref{eq:simp}) and $b'$ its derivative. The contour can be constructed as a ``barbell" figure with three part that we call A, B and C: two lines running along the cut 
in opposite directions (part A) and two circles going around -8 (part B) and 0 (part C) and stopping when they meet the horizontal lines. 
We have monitored the contributions of the three parts for $m\leq 12$ 
and values of $\epsilon \leq 2$ and compared them to reliable numerical values of the derivatives of $M^2_{simpl.}$. 
For small, but not too small values of $\epsilon$,  A and B provide same sign contributions that dominate the integral. As $\epsilon $ decreases, the A and B contributions become large and of opposite 
sign while C becomes negligible. As $\epsilon$ keeps decreasing, the 
absolute values of A and B keep increasing until the accuracy of the 
integrals deteriorate. The situation is illustrated for $m=5$ in Table \ref{tab:int}.  Again, we are facing the challenge of having two large canceling quantities. We have tried to improve this situation by modifying the radius of the circles and the distance between the 
horizontal lines, but the general features seem quite persistent. 
\begin{table}[t]
\begin{tabular}{||c|c|c|c|c||}
\hline
 $\epsilon$ & T =$\di_5(\epsilon)$&A/T&B/T&C/T\\
 \hline
 2.0 & -0.00116908 & -0.166729 & 0.536511 & 0.630219 \\
 1.9 & -0.00202112 & 0.257416 & 0.37977 & 0.362814 \\
 1.8 & -0.00296429 & 0.441387 & 0.320456 & 0.238158 \\
 1.7 & -0.00389598 & 0.527325 & 0.305554 & 0.167121 \\
 1.6 & -0.00465005 & 0.552561 & 0.325353 & 0.122087 \\
 1.5 & -0.0049979 & 0.517248 & 0.390845 & 0.0919072 \\
 1.4 & -0.00467868 & 0.379432 & 0.548854 & 0.0717137 \\
 1.3 & -0.00347855 & -0.0515184 & 0.99067 & 0.0608484\\
 1.2 & -0.0013698 & -2.53523 & 3.45779 & 0.0774331 \\
 1.1 & 0.00131813 & 6.10486 & -5.07936 & -0.0255078 \\
 1.0& 0.00380812 & 3.56912 & -2.56953 & 0.000410349 \\
 0.9 & 0.00507521 & 3.93213 & -2.93389 & 0.00175748 \\
 0.8 & 0.00449056 & 6.30264 & -5.30373 & 0.0010864 \\
 0.7 & 0.00257563 & 16.7544 & -15.7548 & 0.000454331 \\
 0.6 & 0.000812222 & 93.4553 & -92.4554 & 0.000131085\\
 0.5 & 0.0000987024 & 1576.99 & -1576.15 &0.0000229591\\
\hline
\end{tabular}
\caption{\label{tab:int} Values of the contributions A, B and C described in the text in units of the total value T = $\di_5(\epsilon)$ .}
\end{table}

The lesson that can be learned from the second representation is that 
the contributions from small negative coupling, which becomes
the integral around a cut circle around $M^2=-8$ (part B)  after the change of variable, is essential to compensate the large contributions from the cut. 

\section{Conclusions}
In summary, for $2D$ nonlinear $O(N)$ sigma models, the exact discontinuity of the average energy appears to be 
a purely non-perturbative phenomenon. At leading order in the $1/N$ 
expansion the series terminates and no pathological behavior at negative coupling can be inferred from it. It would interesting to see if this feature persists for sub-leading corrections. 

In the large-$N$ and large volume limit, the Fisher's zeros can only 
appear inside a clover shape in the complex $\tc $ plane. It is plausible that as the volume increases, the zeros become dense at the boundary of the clover shape and at the boundary of an infinite set of concentric clover shapes. If this is correct, the zeros closest to the 
real axis in the $b$ plane appear at infinite $Reb$ and $Imb=\pm i/4$.
Numerical studies at finite $N$ and $V$ should clarify the picture 
and could be used as a guide for the search of Fisher's zeros in gauge theories. It would also be interesting to consider the case $D=3$, where there is a rich phase diagram \cite{bmb,david84,Kessler:1985ge} and where we expect the zeros to pinch the real axis in the 
infinite volume limit. 

A simplified form of the discontinuity was approached with dispersive 
methods. An important feature observed was that large contributions 
cancelled. Again, finite volume studies may clarify the mechanism. 
If we extend the reasoning used for one plaquette \cite{npp}, 
non perturbative effects become important at an order $k\sim\beta V$. 
Near this order, the coefficients become different if we integrate a 
perturbative expansion of the density of state from 0 to $DV$ or from 0 
to $\infty$. This understanding is crucial if we want to modify the weak 
coupling expansion in order to include non-perturbative effects. 

We have noticed that the zeros of the $\beta$ function are related to 
the singular points of the mapping $b(M^2)$. 
At finite volume, these singular points have a nonzero imaginary part.
This should be seen as an encouragement to study complexified 
renormalization group flows as also suggested by other work \cite{Polyakov:2004br, Kaplan:2009kr}.

\begin{acknowledgments}
We thank C. Bender, D. Du, H. Sonoda and H. Zou for discussions and encouragement. 
 This work was initially motivated by the workshop 
``New frontiers in large-N gauge theories" where some preliminary results were presented. We thank the organizers and participants 
for stimulating discussions.
The manuscript was written in part while at the Aspen Center for Physics in June 2009. This 
research was supported in part  by the Department of Energy
under Contract No. FG02-91ER40664.
\end{acknowledgments}


\end{document}